\documentstyle[12pt,preprint]{aastex}

\input{psfig}

\begin{document}

\title{Stellar models of evolved secondaries in CVs.}

\author{N. Andronov, M. H. Pinsonneault}
\affil{Ohio State University, Department of Astronomy, Columbus, OH 43210}
\affil{E-mail: andronov, pinsono@astronomy.ohio-state.edu}

\begin{abstract}

In this paper we study the impact of chemically evolved secondaries on CV evolution. 
We find that when evolved secondaries are included a spread in the secondary 
mass-orbital period plane comparable to that seen in the data is produced for 
either the saturated prescription for magnetic braking
or the unsaturated model commonly used for CVs. We argue that in order 
to explain this spread a considerable fraction of all CVs should have evolved 
stars as the secondaries. The evolved stars become fully 
convective at lower orbital periods. Therefore, even if there was an abrupt decrease 
in magnetic braking for fully convective stars (contrary to open cluster data) it would 
not be expected to produce a sharp break in the period distribution for CVs. 
We also explore recent proposed revisions to the angular momentum loss rate for single
stars, and find that only modest increases over the saturated prescription are
consistent with the overall observed spindown pattern. We compare predictions of our 
models with diagnostics of the mass accretion rate in WDs and find results intermediate 
between the saturated and the older braking prescription. Taken together these 
suggest that the angular momentum loss rate may be higher in CV secondaries than 
in single stars of the same rotation
period, but is still significantly lower than in the traditional model. Alternative 
explanations for the CV period gap are discussed. 
\end{abstract}

\section{Introduction}
Cataclysmic variables (CVs) are mass transferring 
close binaries (Patterson 1984; Warner 1995). The primary, or 
accreting star, is a white dwarf. The secondary is a low mass main sequence star which 
overfills its Roche Lobe and transfers matter onto the primary.

The evolution of cataclysmic variables is driven by two major ingredients: the angular 
momentum loss and the response of the secondary star to the mass loss. In the standard 
model of CV evolution (Rappaport, Verbunt \& Joss 1983) angular 
momentum loss is assumed to be extremely efficient until the secondary becomes 
fully convective, at which point it ceases.  This implies a rapid evolutionary 
timescale, numerous very low mass CVs, and a high average mass accretion rate; the
secondary stars in CVs lose mass on a timescale shorter than the Kelvin-Helmoholtz 
timescale and are out of thermal balance. In this ``standard'' model the observed 
period gap is explained 
by a reduction in the angular momentum loss rate at the onset of full 
convection, which causes the secondaries to relax to a smaller radius and fall 
out of contact. Such a model was consistent with the stellar data at the 
time it was proposed. 

In an earlier paper (Andronov, Pinsonneault \& Sills 2003, hereafter APS) we showed that 
recent stellar data on angular momentum loss is inconsistent with the standard CV model. 
Fully convective single stars are observed to experience angular momentum loss, and the 
rate of angular momentum loss saturates at a much lower level than that given by a simple 
extrapolation of the solar rate to the periods of interest for CVs.  The inferred timescale 
of CV evolution is much longer. In addition, the secondary evolves during the pre-CV phase. 
These effects raise the possibility that the secondary 
stars in CVs could be significantly evolved prior to the CV phase. In APS we explored 
some simple models with chemically evolved secondaries; nuclear evolution during the CV 
phase was neglected. In this paper we explore the effects of the evolutionary state of 
the secondary in more detail.

We therefore begin in section 1.1 with a discussion of our angular momentum loss 
rates, which motivate the exploration of evolved 
CV secondaries in section 1.2. The important question of the origin of the period gap is
discussed in section 1.3. In section 2 we briefly describe our models and method.
We begin our results by presenting models with different evolutionary states
and angular momentum loss rates in section 3. We also examine the secondary mass - 
orbital period relation. In section 4 we discuss tests of the braking law for 
single stars and the mass accretion rate in CVs.
Section 5 is devoted to a discussion of various classes of solutions for
the origin of the period gap. We find two new possibilities that deserve further study: 
tidally-induced mixing of evolved secondaries and non-equilibrium CV period 
distribution tied to the star formation rate in the Galaxy.

\subsection{Angular Momentum Loss in CVs}

There are two mechanisms for angular momentum loss in CVs. The first is gravitational 
radiation (Landau \& Lifshitz 1962), which is effective only for short 
orbital periods (Patterson 1984) and therefore can not be the 
only mechanism responsible for angular momentum loss.
The second is angular momentum carried away by a stellar wind magnetically coupled to 
the surface of the secondary. In tidally-locked binaries (such as CVs) this momentum 
is removed from the total orbital momentum of the system, causing a secular decrease 
of the orbital period.

There has been significant progress in the modeling of angular momentum loss from 
a magnetized stellar wind (see Appendix A of Keppens, MacGregor \& Charbonneau 1995). 
In early models of the solar wind (Weber \& Davis 1967) solutions of the MHD equations 
were obtained in the equatorial plane.  More recent calculations solve the full MHD 
equations in three dimensions. One prediction of the early models was that at low rotation 
rates the mean surface magnetic field strength would scale as the rotation rate $\omega$ and 
the angular momentum loss rate would scale as $B^2 \omega$ or $\omega^3$. Such a scaling 
was consistent with early data on the spindown of stars (Skumanich 1972). However, at high 
rotation rates the angular momentum loss rate grows at a slower rate even with the same 
underlying dynamo model (Keppens et al. 1995), and there is the possibility that the 
strength of the magnetic field would saturate when the surface became filled with spots. 
Proxies for the magnetic field strength, such as X-ray fluxes and chromospheric activity 
indicators, exhibit a mass-dependent saturation at high rotation rates 
(Stauffer et al. 1997). 

Both the morphology of stellar magnetic field and the properties of stellar coronae 
are difficult to infer and complicate estimates of angular momentum loss. 
There are also uncertainties related to the nature of the dynamo mechanism itself. The most 
popular model for the solar cycle is the interface dynamo (Parker 1993; MacGregor \& 
Charbonneau 1997; Charbonneau \& MacGregor 1997; Montesinos et al. 2001). In this 
model the toroidal magnetic fields are generated in a shear layer below a 
surface convection zone from a poloidal field 
(the $\omega$ component of an $\alpha - \omega $ dynamo), while the poloidal field is 
regenerated from the toroidal field in a nearby but different layer (the $\alpha$ component.) 
Such models are more successful in reproducing the properties of the solar cycle than 
the classical thin-shell dynamo where both effects take place in the same region. 
Montesinos et al. (2001) extended such models to other stars.

However, neither a thin shell dynamo nor an interface dynamo would operate in a fully 
convective star. Instead the generation of magnetic fields would require a distributed 
or turbulent dynamo that would be less effective (see Lanza et al. 2000 for a 
discussion of potential implications of the interface 
dynamo for CVs). However, it is not clear that the theoretical models have 
sufficient predictive power to infer the absolute efficiency of different dynamo mechanisms. 
This induces a further model dependence in the angular momentum loss rates as a 
function of mass predicted by theory.

We have therefore chosen to use open cluster stars with a range of mass and age 
to empirically measure the angular momentum loss rates as a function of mass and 
rotation rate. The overall trends of this prescription for magnetic braking are:\\ 
1. It saturates at a level that scales inversely with the convective 
overturn timescale for masses greater than 0.6 $M_{\odot}$ (see Krishnamurthi et al. 1997). 
This suggests consistency with either an interface or thin shell dynamo. It is possible
to construct models of rapid rotators that do not have a saturation threshold for magnetic
activity (Solanki, Motamen \& Keppens 1997), but the torque is comparable to that obtained
with saturated models.\\  
2. It drops at a faster rate for lower mass stars (see Sills, Pinsonneault \& Terndrup 2000), 
but with no abrupt change at the fully convective boundary; 
this suggests that the transition to a different dynamo mechanism is gradual 
rather than abrupt.  

In APS we used these empirical measures to infer the loss rates for 
secondary stars in CVs. We discuss tests of the braking law in section 4.

\subsection{Physics of CV secondary stars.}

The less efficient SPT magnetic braking prescription implies a slower rate of period 
decrease for CVs than the RVJ prescription does. This 
can be tied to the mass loss rate through the response of a star to mass loss. The net effect 
is a lower time-averaged mass accretion rate. 
The typical lifetime of a CV with an initial secondary mass of order $1 M_{\odot}$ or more
becomes comparable to its main sequence lifetime. As a result, it is not clear that 
nuclear evolution during the CV phase can be completely neglected.

In addition, there are two major effects that could permit secondary stars to begin the CV 
phase with significant internal nuclear processing. The primary must exhaust its
central hydrogen prior to a common envelope phase. The secondary will also evolve 
which is an important effect prior to common envelope phase only for secondary 
stars similar in mass to the primary. However, there is also a potential time delay between
the common envelope phase and the onset of the CV phase. 
If the output of the CE phase is a relatively wide binary (compared to CVs) with 
an orbital period of a few days, then the time for the system to reach contact may be
significant and could result in appreciable nuclear evolution
of secondaries prior to the CV phase. Furthermore, the low inferred loss rates for very 
low mass secondaries may favor higher mass secondaries with shorter nuclear timescales. 
Therefore, CVs with evolved secondaries may contribute to CV populations. Hence it seems 
reasonable to study how the evolutionary state of the secondary affects the properties of CVs.

From an observational standpoint, the evidence for a considerable fraction 
of evolved systems comes from the spread of the mass of secondaries at 
a given orbital period (Smith \& Dhillon 1998). This spread is naturally explained 
if we assume that some secondaries experienced significant nuclear evolution prior to
the CV phase (APS for example). Baraffe \& Kolb (2000) came to the same conclusion
considering the spectral type - orbital period relation in the data by Beuermann at al. (1998). 
Most theoretical work has assumed that secondaries are chemically 
unevolved.\footnote{Recently Podsiadlowski, Han \& Rappaport (2003) have 
examined this issue in the framework  of the old angular momentum loss model. We discuss 
their results in section 3.2.} 

In sections 2.4 and 2.5 we discuss two 
mechanisms which can change mass-radius relationship for the secondary star in CV, but
are not included in recent models. These mechanisms are related to existence of spots
on the surface of rapidly rotating stars and extra mixing associated with 
tidal deformation of a star.

\subsection{The origin of the period gap}

The study of cataclysmic variables has a rich history.  The theoretical framework that 
is traditionally employed has had some success in explaining many of the global features 
of the observed population of CVs. However, there are some key assumptions in the 
``standard'' model, most notably concerning angular momentum loss, that are seriously 
inconsistent with both the observed spindown of young low mass stars and theoretical 
developments in our understanding of stellar winds. A critical re-examination of the 
physical effects responsible for some of the major properties of CVs is therefore clearly 
warranted.

For example, the presence of only a few CVs with periods between 2 and 3 hours 
(the period gap) is certainly a critical component of the CV phenomenon. If the
CV distribution is in equilibrium, then the most natural interpretation is that 
there is a physical effect that causes stars to shrink at a characteristic orbital
period of 3 hours. The ``standard'' model uses a sudden drop in the efficiency
of magnetic braking for fully convective stars to achieve this. Such a phase transition 
would produce a sharp break in the spindown properties of 
single stars which is not seen.

However, there are other physical mechanisms that could potential
cause changes in the mass-radius relationship near the fully convective boundary. 
As a result, it is entirely possible that the ``standard'' model is on the right 
track (by solving the period gap with a change in the mass-radius relationship),
but that it is not using the proper underlying physical mechanism. Alternately, 
the period gap could arise from the presence of either more than one underlying 
population or from the distribution of CV periods not being in equilibrium.  In 
our view all of these possibilities should be explored, and a better physical 
picture of the evolution of cataclysmic variables will be obtained. In section 5 we 
discuss different possible scenarios for a formation of the period gap.

\section{Physics}

In this section we describe the physics of CVs relevant to our
calculations. All capital letters in this section and farther
denote quantities in cgs units while all small letters express
quantities in dimensionless units relative to the sun.

The evolution of a close binary which eventually forms a CV can be described in 4 stages.\\
1. Main sequence evolution. The more massive (primary) star leaves the main
sequence first and expands on the red giant branch. 
A common envelope (CE) system forms when unstable mass accretion onto the secondary sets in
(De Kool 1990; Iben \& Livio 1993).\\
2. CE stage of evolution (e.g. De Kool 1990). This phase is short 
($\approx 10^4$ years); during this stage the secondary spirals in towards 
the primary and the gravitational potential energy is absorbed by an envelope which 
is subsequently ejected. After the envelope is ejected, the system consists of a 
white dwarf and a main sequence secondary star which does not necessarily overfill 
its Roche Lobe but might be close to it. The final separation of the main sequence star
and the white dwarf should depend on the initial mass ratio of the stars in a binary
and their initial separation. However, the current understanding of this phase of evolution
is not adequate to predict the outcome of this phase (De Kool 1990). The only 
conclusion we can make is that at least some systems after this phase are close 
enough (with orbital periods below $\approx$ 5 days) to form a CV eventually.\\ 
3. Post-CE and Pre-CV evolution. During this phase stars in the binary get closer due 
to angular momentum loss from the system until the secondary overfills its Roche Lobe.\\
4. CV evolution. The outcome of the first 3 phases defines the starting point of 
the CV phase. The relevant ingredients are the masses of the white dwarf and the 
secondary ($m_{wd}$ and $m_2$) and the evolutionary state of the secondary which 
determines the $M-R$ relationship. We characterize the evolutionary state by its 
central hydrogen abundance ($X_c$).

For the models in this section we used a primary mass of white dwarf $0.85 M_{\odot}$
the average measured mass of CV primaries (Patterson 1984).\footnote{APS used a smaller mass 
of $0.62 M_{\odot}$, typical for field WDs.}
We keep it constant during the evolution. The hydrogen rich material 
accumulated on the surface of the white dwarf is assumed to be lost during nova 
outbursts (Warner 1995). The timestep used in our models is considerably larger 
than the nova duty cycle, which insures that we average 
the behavior of the system over many such cycles.

We assume that the evolution of the secondary is unaffected by the companion 
until it becomes a CV. We therefore generate a grid of secondaries described by their 
mass and central hydrogen abundance. These two parameters are initial conditions
for phase 4 and uniquely describe the behavior of the system at the onset of the
mass transfer and along the CV stage of evolution.

\subsection{Stellar model}

We used the Yale Rotating Stellar Evolution Code (YREC) as described in APS for the standard 
model physics of our code (equation of state, nuclear reaction rates, opacities, boundary 
conditions, and the heavy element mixture). We examined models with solar heavy element 
abundance with a range of initial masses. Once the secondaries overfill their Roche 
lobe we include mass loss in the stellar interiors calculations as follows.

In order to account for mass loss, we remove mass from the outer 
convective region given the mass accretion rate and timestep by simply decreasing 
the mass spacing uniformly. This corresponds to the assumption that the specific 
entropy of the convection is unaffected by the mass loss. We then allow the model 
to relax and evolve. The mass accretion rate is derived to preserve conservation 
of angular momentum as described in the following section.
  
\subsection{Angular momentum and its loss}

The orbital angular momentum of the system is given by:
\begin{equation}
J=M_{\odot}^{5/3}G^{2/3}m_1 m_2 m^{-1/3}\omega^{-1/3}
\end{equation}
while the angular momentum loss rate consists of two terms:
\begin{equation}
\left(\frac{dJ}{dt}\right)=
\left(\frac{dJ}{dt}\right)_{grav}+
\left(\frac{dJ}{dt}\right)_{wind}
\end{equation}
The first term is angular momentum loss due to gravitational 
radiation(Landau \& Lifshitz, 1962)
\begin{equation}
\left(
\frac{dJ}{dt}
\right)_{grav}
=-\frac{32}{5} \frac{G^{7/2}}{c^5} a^{-7/2} m_1^2 m_2^2 \sqrt{m}
M_{\odot}^{9/2}
\end{equation}
and the second is the magnetic braking term $\left(\frac{dJ}{dt}\right)_{wind}$ from a 
solar-like wind coming from the secondary.
Here $m_1$ , $m_2$ , $m$ are the white dwarf mass, secondary mass,
and total mass respectively, $a$ is the separation between the stars,
$\omega$ is orbital speed.

For magnetic braking we use two different empirical prescriptions;\\ 
a)The Rappaport, Verbunt \& Joss model (1983) based on early stellar spindown data, 
which is assumed to operate only when the model is not fully convective.\\
b)The  more recent empirical rule obtained from spindown of stars in young 
open clusters (Sills, Pinsonneault \& Terndrup 2000):
\begin{equation}
\left(\frac{dJ}{dt}\right)_{wind} = -K_w \cdot \sqrt{\frac{r}{m}}\cdot
\left\{
\begin{array}{ll}
\omega^3                & {\rm for} \; \omega \leq \omega_{crit} \\
\omega \omega^2_{crit}  & {\rm for} \; \omega > \omega_{crit}
\end{array}
\right.
\end{equation}
where $\omega_{crit}$ is a critical frequency at which generation of magnetic 
field saturates. Above this frequency the angular momentum loss becomes 
linear (in $\omega$) instead of the third power. The constant $K_w=2.7\cdot 10^{47}$g cm s.

In the saturated braking model, we assume that the critical angular 
speed (which is set by a maximum magnetic field which the star 
is able to generate) is a unique function of the effective temperature 
of the star (the upper limit on the magnetic field is set by the effective 
temperature alone). Because the convection zone depth is primarily a function of 
effective temperature, this is not an unreasonable approximation. Evolved stars would 
have larger radii and longer convective overturn timescales than single stars of 
the same mass. 
This implies that the actual angular momentum loss rates for evolved secondaries would be 
somewhat lower than the value inferred for single stars. We include this effect in the 
first approximation, calculating the saturation frequency as a function of effective 
temperature of a star but neglecting any dependence of $\omega_{crit}$ on the 
star's surface gravity.
We therefore infer the angular momentum loss for a star 
by interpolation in the $\omega_{crit} - T_{eff}$ relation for young open clusters
presented in Table 1 of APS. In this way we are able to take into account the change of 
magnetic braking for models of secondary stars with different evolutionary states.

\subsection{Marginal contact}
We use the Eggleton (1983) approximation for the roche lobe radius.
In the 'marginal contact' assumption, it defines the radius of the spherical 
stellar model which overfills its Roche Lobe:
\begin{equation}
R_2 = R_L = a\cdot\frac
{0.49 \left(\frac{m_2}{m_1}\right)^{2/3}}
{0.6\left(\frac{m_2}{m_1}\right)^{2/3}+\ln{\left[1+
\left(\frac{m_2}{m_1} \right)^{1/3}\right]}}
\end{equation}

Given the equatorial radius of the 
model and effective temperature provided by the YREC code, we derive the mass 
accretion rate (as described in APS): it is calculated 
by requiring that equations 1,2, and 5 be consistent with each other. Mass loss rate 
are passed to the stellar code which would subtract the amount of mass from the convective 
envelope given the mass loss rate and timestep, solve for the new structure 
and evolve the model.

\subsection{Starspots.}
We used a model similar to the one used by Spruit \& Weiss (1986). The spots are assumed
to be completely black, cylindrical, extending into the interior to a depth of 10 pressure 
scale heights.\footnote{The results of the model are insensitive to the depth of the spots, 
as long as they are considerably deeper than superadiabatic region.} In our models
50\% of the surface area is covered by spots. The convective energy transport is completely 
suppressed within a spot. Following Spruit \& Weiis (1986), we assume that
convective energy transport through a spherical shell of radius $R$ covered by spots becomes;
$$
L_r=4\pi r^2 F_0(1-f_s)
$$
where $F_0$ is the local energy flux in the areas outside spots and $f_s$ is the fractional 
area covered by spots.

The results of this model are described in section 5.1.2.
In our model we completely ignore the radiative energy transport through the spot, so
that we get a maximum possible effect on stellar parameters; more realistic models of starspots
would have smaller effect on the radius and luminosity of a star.

\subsection{Rotational mixing.}
Rotation can induce mixing through meridional circulation and other 
hydrodynamical mechanisms included in single star models. 
The possible mixing of elements associated with the tidal deformation 
of a star has not been taken into account in models of CV secondaries. To estimate the
importance of such effect we compare the thermal timescale
of the radiative core to the timescales of mixing. For qualitative estimate
of this effect, we assume that large scale currents caused by deformation
have a characteristic timescale defined by $\tau_{KH}/d$, where
$$
\tau_{KH}\approx\frac{GM^2}{RL}
$$
is the thermal (Kelvin-Helmholtz) timescale and $d$ is a dimensionless 
parameter describing the departure from spherical symmetry.
In the case of pure rotation, the departure from spherical symmetry is defined
as the ratio of centripetal acceleration for a point on the equator to the gravity at this point,
$$
d\approx\frac{\omega^2 R^3}{GM}
$$
If we are dealing with stars in a close binary system, the tidal force from the companion
can be significant. For a qualitative order of magnitude estimate of this effect
we adopt the Roche model as a loose representation of equipotential surfaces of star and 
define the departure from spherical symmetry as
$$
d\approx\frac{\left.\psi(m_1)\right|_{a+R_{rad}}
-\left.\psi(m_1)\right|_{a-R_{rad}}}
{\left.\psi(m_2)\right|_{R_{rad}}}
$$
where $\psi(m_1)$ and $\psi(m_2)$ are the gravitational potentials of the white dwarf
and the secondary, $a$ is the distance between white dwarf and the center of the secondary, 
$R_{rad}$ is the size of the radiative core of the secondary. 
 
The results can be found in section 5.1.2.  

\section{Results} 
The overall framework of our models discussed in section 2 is similar to that in APS.
Our approach is different from APS in the use of full stellar models to calculate 
a radius as a function of time, instead of assuming a mass-radius relationship. This allows
us to calculate models with arbitrary initial evolutionary state and abundance, as well as models
in which mass loss is sufficiently large to drive secondaries out of thermal balance. 
Comparison of such models with observable quantities can help constrain the physics 
and population of CVs. These observable quantities include:\\
1. The secondary mass - orbital period relationship. Within the framework described above, 
the M-P relation might help constrain the evolutionary state and M-R relationship of CV secondaries.
Given the insensitivity to the mass of the primary, we can use this to learn about the pre-CV phase
of evolution.\\
2. The mass accretion rate as a function of mass of the secondary or orbital period
might give clues about the angular momentum loss rate and the response of the 
secondary to mass loss (which is a function of the evolutionary and thermal 
state of the secondary). However, it is difficult to make apple to apple 
comparison between derived and observed accretion rates.
The problem lies in the timescale associated with accretion. While the luminosity of CVs
and therefore the instantaneous accretion rate are determined by the physics of accretion disks,
model accretion rate is averaged over long periods of time (~$10^7$) years.
This means that any comparison of derived mass accretion rate with observed one should be taken
with extreme care. We chose the recent observed data by Townsley \&
Bildsten (2003b). They derive mass accretion rates by measuring white dwarf temperatures 
and calculating the impact of accretion on the thermal state of a white dwarf. 
The mass accretion rate measured this way is effectively the time averaged rate
over a timescale of about $10^3$ years. 

\subsection{Saturated magnetic braking.}
As discussed in APS, modern studies of the angular momentum evolution of low mass stars,
require a much milder angular momentum loss rate than that typically used for CVs. Angular 
momentum loss rate of fast rotators saturates at some rate, above this threshold it scales
linearly with $\omega$ instead ot $\omega^3$ dependence. This has dramatic effects
on the properties of CVs. As it was shown in the previous paper, angular momentum loss rates are
not sufficiently large to drive mass accretion high enough which would cause secondary 
move from normal thermal equilibrium to a new one and puff up. The secondary is then at 
normal thermal balance during all its lifetimes as a secondary in CV. 

The evolutionary tracks for our saturated models are shown in figure 1.
Solid lines represent the models starting from the ZAMS with different initial secondary masses. 
The thick solid line is a model with an initial secondary mass of $0.9 M_{\odot}$; dashed 
lines represent models with the same initial mass
but with different initial evolutionary status (central hydrogen content).

\begin{figure}[t]
\plotone{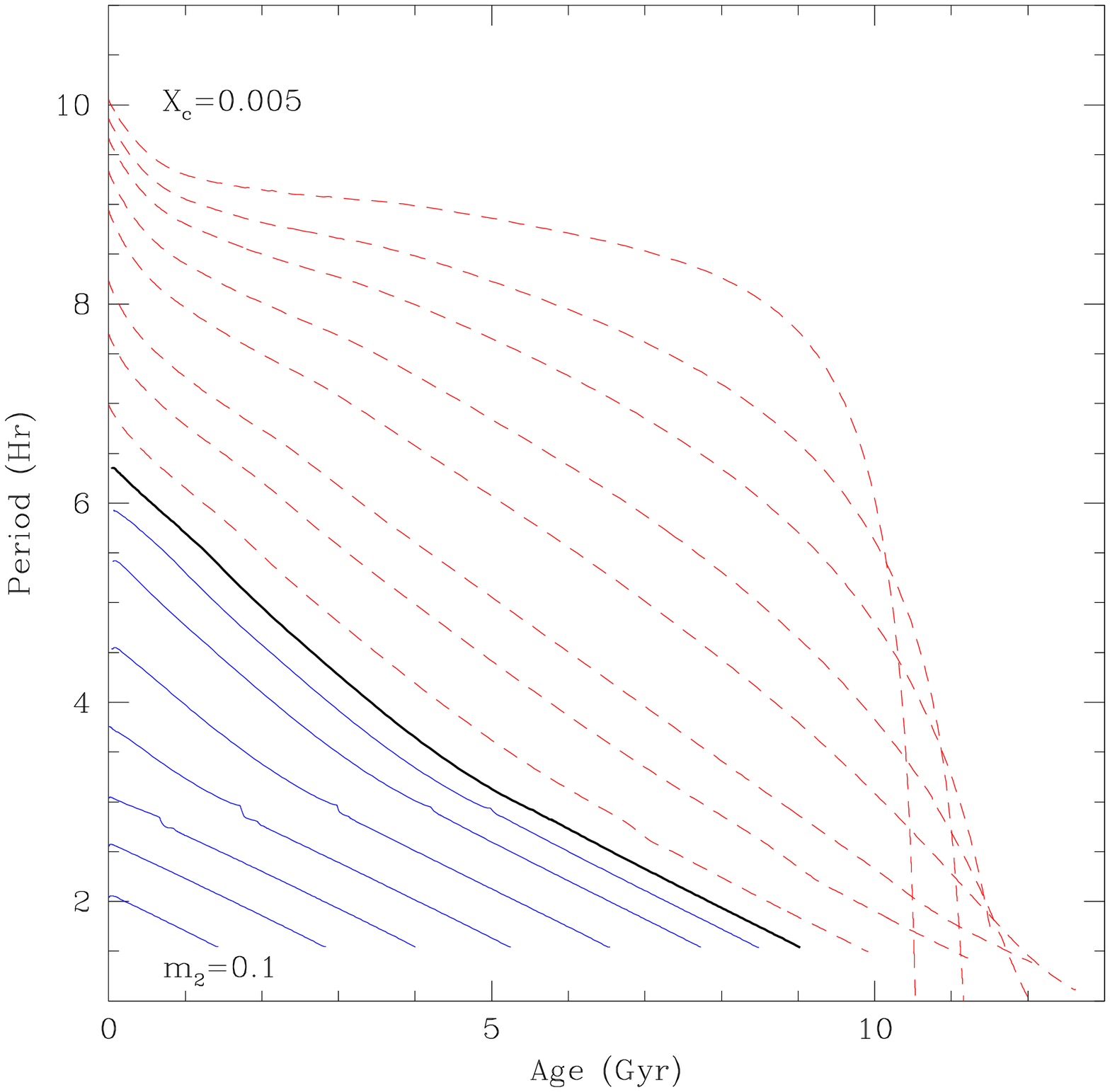}
\caption{Orbital period as function of time for the saturated braking law. The 
thick solid line is the unevolved model with an initial mass of 0.9 $M_\odot$. The dashed 
lines are evolved models, with initial
central hydrogen abundance $X_c$ of 0.50,0.30,0.20,0.10,
0.05,0.02,0.01,0.005.
Thin solid lines are the models starting from ZAMS, but 
with different initial masses (0.1 $M_{\odot}$ increments).}
\end{figure}

For low mass tracks there is a noticeable bump when the system becomes fully convective 
even for the lower saturated angular momentum loss rates. 
This happens at an orbital period of about 3 hours; the system re-establishes contact 
at a slightly lower period (around 2.75 - 2.80 hours). This is related to the sudden mixing of 
$He^3$ when a star becomes fully convective. Originally this $He^3$ mixing was proposed to be
a cause of the period gap (D'Antona and Mazzitelli, 1982).
Consistent with prior results (e.g. McDermott \& Taam, 1989), we find that 
the width of this feature and timescale are certainly not wide enough and long 
enough to explain the CV period gap between 2 and 3 hours. 
We, therefore, conclude that a small period gap would 
exist even in the case when the mass transfer rates are insufficient to drive the secondary out 
of thermal equillibrium, but only when most CV secondaries have a central helium 
abundance close to the ZAMS value. The evolved tracks do not follow a unique 
evolutionary path $P(m_2)$ in the sense in which unevolved ones do.

\subsection{Disrupted unsaturated magnetic braking}

In this section we summarize the results of the models with magnetic braking 
in the form proposed by Rappaport, Verbunt \& Joss (1983). 
In these models the mass accretion rate is so high that the timescale for mass loss 
becomes shorter than the Kelvin-Helmholtz timescale. The model is out of thermal 
balance and becomes larger for a given mass than a ZAMS star. When the 
star becomes fully convective, the torque from a magnetic wind is shut down. 
The star then detaches and relaxes to its normal radius in 
a thermal timescale. When contact is reestablished, the driving 
mechanism of CV is angular momentum loss by gravitational radiation alone. It is not 
efficient enough to drive the star out of thermal balance, and the subsequent evolution 
proceeds with a considerably slower pace than was when magnetic braking was 
operating. We ran the same set of models which we did for the unsaturated prescription. 

\begin{figure}[t]
\plotone{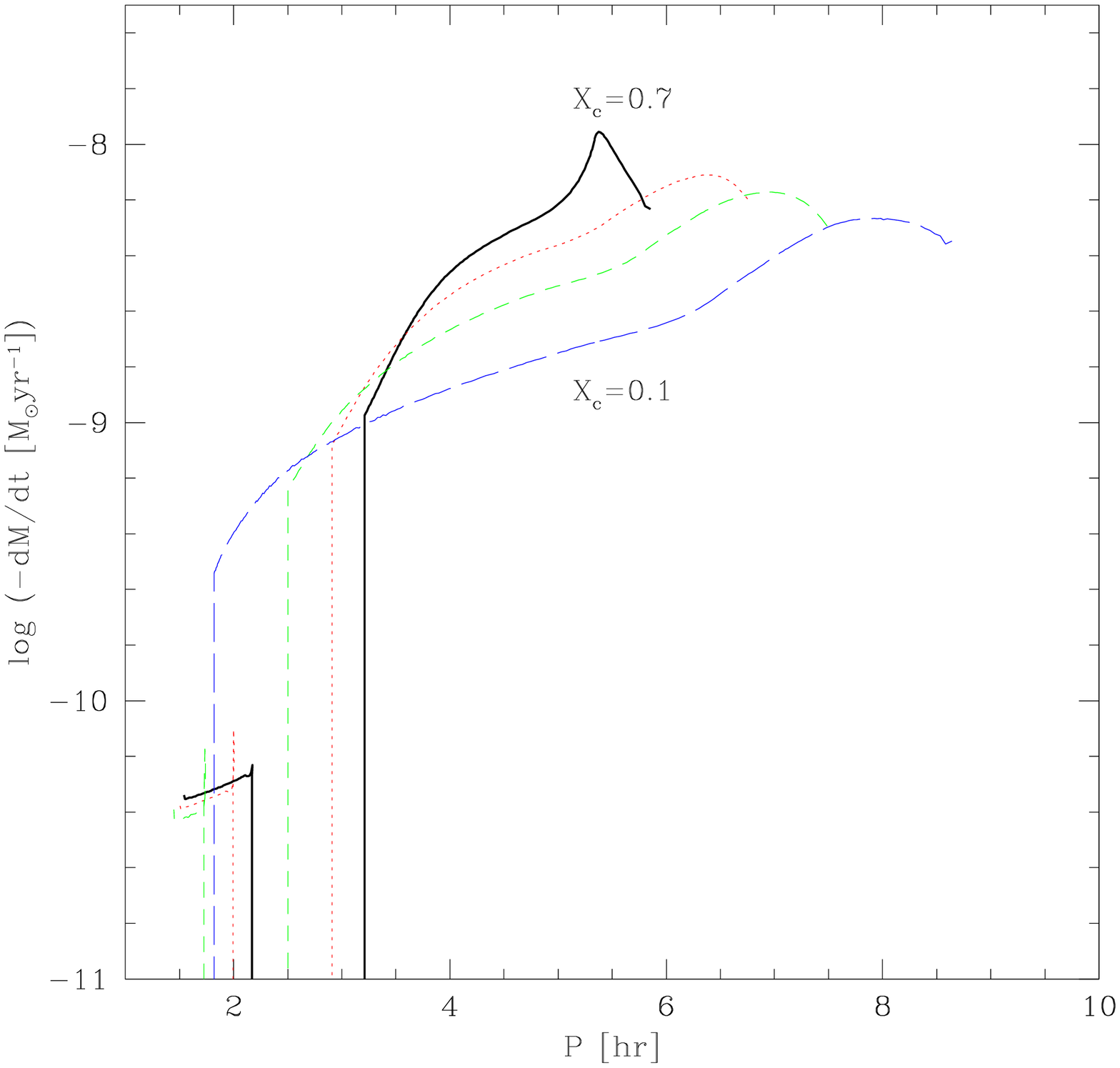}
\caption{Mass accretion rate as a function of period for the unsaturated prescription. 
The thick solid line is an unevolved model. The thin dashed lines are the models 
pre-evolved to $X_c$ = 0.50,0.30,0.10}
\end{figure}

The mass loss rate for models with RVJ braking are shown in figure 3. ZAMS models roughly
reproduce the Period Gap. However, more evolved ones fail to do it. They become 
fully convective at periods lower than 3 hours and become 'active' again 
(and therefore 'visible') at periods shorter than 2 hours. In addition the period gap becomes 
smaller and smaller at lower $X_c$, becoming quite narrow and completely vanishing 
for models with an initial central hydrogen abundance below 0.2. Figure 4 shows 
the boundaries of the gap as a function of central hydrogen abundance.

\begin{figure}[t]
\plotone{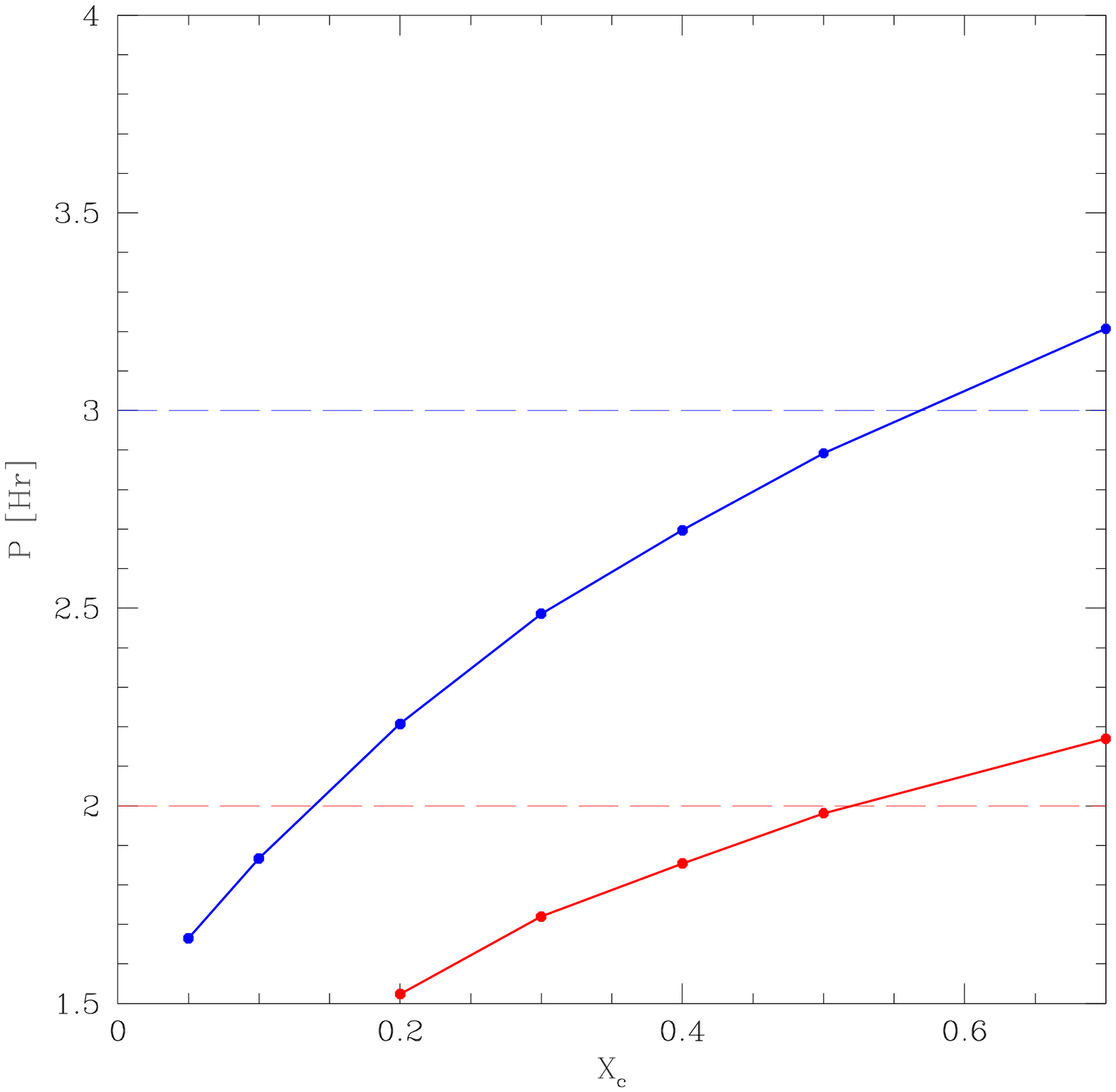}
\caption{Boundaries of the period gap in the disrupted magnetic braking model. Thick
solid lines are the upper and lower boundaries as a function of the central hydrogen abundance
at the onset of the CV phase. Thin dashed lines show the observational location of the gap.}
\end{figure}

This might be a significant problem for the 'disrupted magnetic braking model' 
as an explanation of the period gap. The observed mass-period relation indicates 
that evolved secondaries constitute a considerable fraction of all 
CVs (see Patterson 1984, Smith and Dhillon 1998). From a theoretical view, evolved 
models are also very well motivated; the secondary had to evolve during the MS 
evolution of pre-CE binary, and during the pre-CV phase. The gap would be washed 
out by the population of evolved systems, which have a gap at lower periods, 
for a more narrow range of periods, or do not have it at all.

Recently Podsiadlowski, Han \& Rappaport (2003) have examined this issue.
Consistent with what we report here, they found that with the old angular
momentum loss model evolved secondaries would produce a narrower gap shifted to
shorter periods. They then relied on population synthesis models to claim that the period
gap would nonetheless persist, largely because the evolved secondaries only dominate
the population at long periods. It is difficult to quantify the errors in such synthesis
models, which depend on a series of ingredients that are difficult to test.  Quite apart from
the uncertainty in the braking law, the outcome of the common envelope phase will have a profound
impact on such synthesis models and our knowledge of this process is limited. For
this reason we believe that the best approach is to identify regimes where 
the various physical effects can be separated; the spread in mass at fixed 
period just above the period gap does seem to provide such a diagnostic tool.

\subsection{The Secondary Mass-Orbital Period relationship}

The secondary mass-orbital period relationships for our models are shown in 
figures 5 and 6. The data are taken from Smith and Dhillon (1998).

\begin{figure}[t]
\plotone{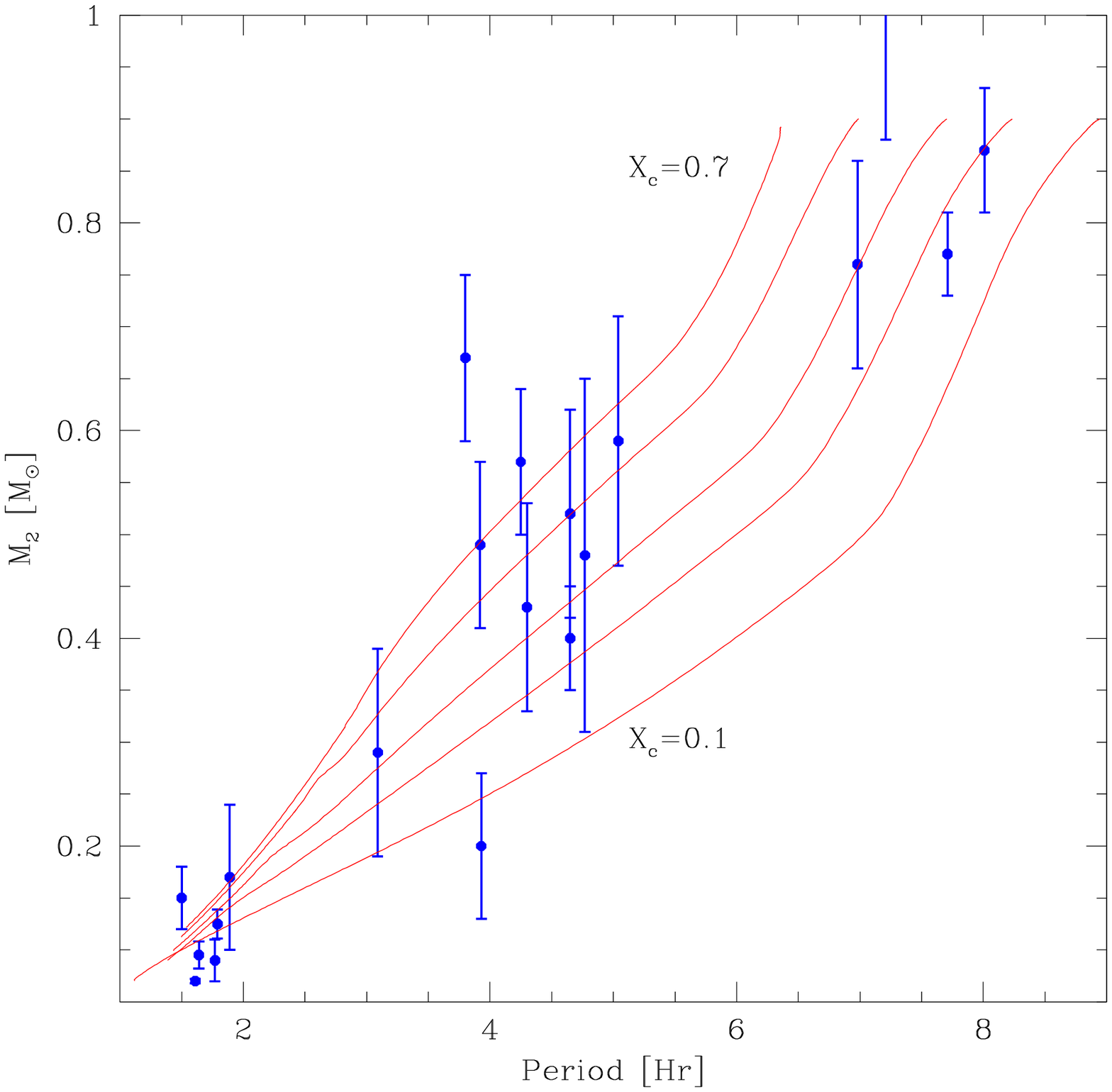}
\caption{The mass-period relation for models with saturated angular momentum loss law. 
Data points are taken from 
Smith and Dhillon (1998). Lines represent models with initial secondary mass of 
0.9 $M_{\odot}$, pre-evolved to $X_c$ = 0.70,0.50,0.30,0.20,0.10.}
\end{figure}

\begin{figure}[t]
\plotone{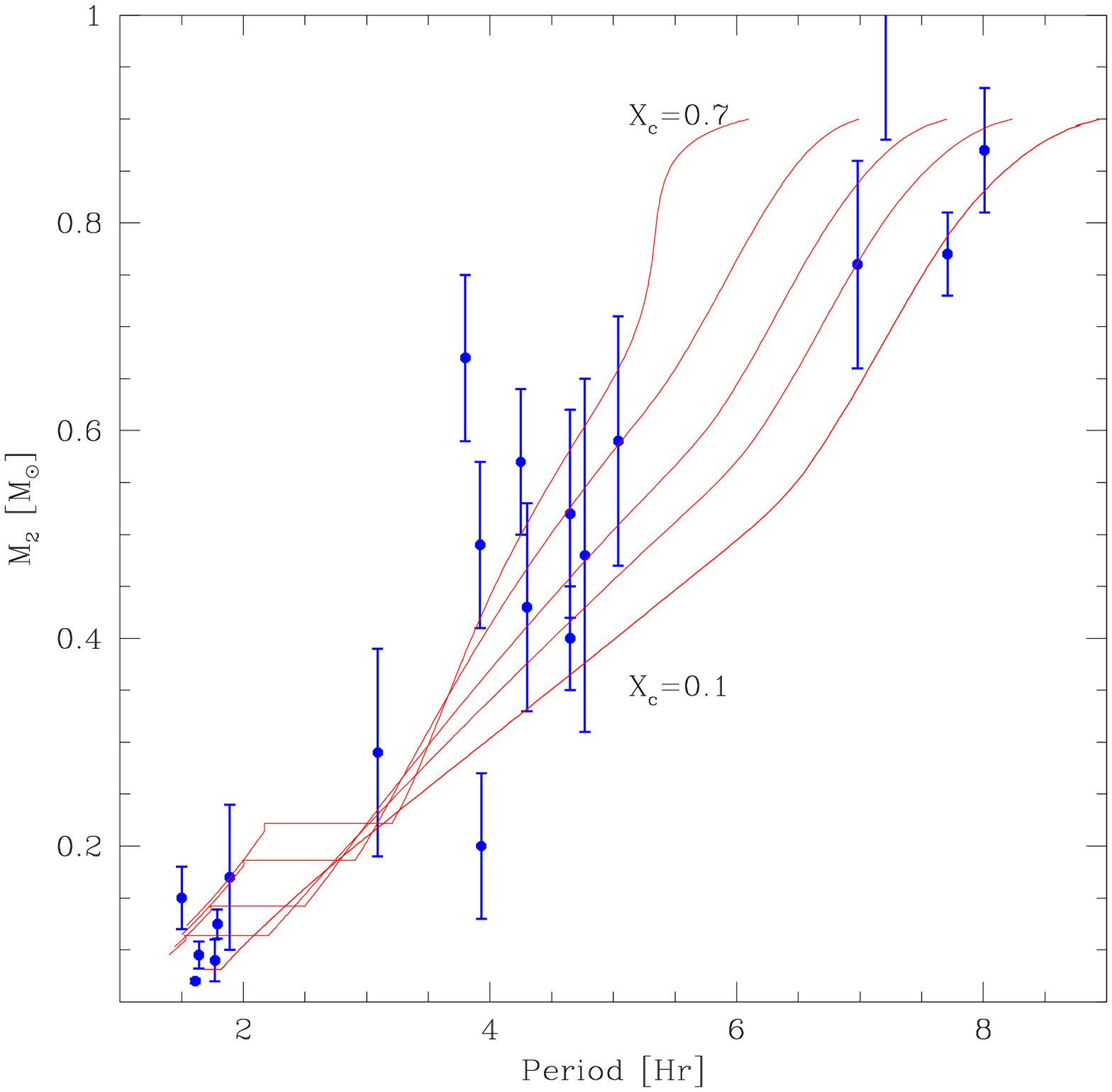}
\caption{The mass-period relation for models with unsaturated angular momentum 
loss law. Data points are taken from Smith and Dhillon (1998). Lines represent models with initial 
secondary mass of 0.9 $M_{\odot}$, pre-evolved to $X_c$ = 0.70,0.50,0.30,0.20,0.10.}
\end{figure}

As has been shown before (for example Baraffe \& Kolb 2000, APS) the models with 
different evolutionary state are able to reproduce the 
observational spread in the mass of the secondary for a given orbital period. Here we compare 
results obtained for 2 magnetic braking prescriptions with full calculation 
of stellar structure and therefore nuclear burning along CV phase. 
The results are shown in figure 5 (saturated braking) and figure 6 (unsaturated braking).

The main conclusion is that both prescriptions can generate a spread in $M$
at fixed $P$ that roughly matches that seen in the data. 
The existence of such a spread is therefore primarily a diagnostic 
of a mix of evolutionary states rather than a test of the braking law.
However, there is a difference in the amount of spread for periods slightly above the period gap 
(3 - 4 hours), where the saturated prescription produces a larger range than 
the unsaturated model does. With improved statistics, the mass distribution in this period range
could be a test of the empirical braking laws. 

\section{Uncertainties of the model}
Our models are based on the application of angular momentum loss inferred for
single stars to the case of close binaries. Two major questions arise here:\\
1. How precisely is the angular momentum loss rate constrained for the single stars?\\
2. Is an empirical angular momentum loss rate derived for the single stars 
applicable to the case of close binaries?\\
We address these two questions in this section. We will demonstrate that angular 
momentum loss rates for single rapidly rotating stars are constrained relatively well. 
The answer to the second question is less clear. There is no strong decisive 
theoretical reasoning or observational evidence that a star in a binary would suffer 
similar (or different) angular momentum loss as a single star rotating at the same rate.

\subsection{Magnetic braking in single stars revisited}
Historically an angular momentum loss rate was derived using an average rotational
velocities ($Vsin(i)$) of stars (or rather distribution of rotations)in populations of 
different ages. An angular momentum loss rate is assumed to have some
functional dependence of mass, radius, effective temperature and rotational rate
of a star, which is motivated by theoretical considerations and then calibrated
using rotational data. Besides the observational errors, there are two main ingredients 
which determine the precision with which the angular momentum loss rate is constrained:\\
1. The assumed initial conditions.\\
2. Ages of the stellar populations which are used.\\
There are other ways of testing models of magnetized stellar winds.
The X-ray luminosity of a star is one of the fingerprints of magnetic activity,
and therefore it can be used to constrain the properties of stellar magnetic fields.
The observed saturation of $L_x/L_{bol}$ at high rotation rates has been used 
as evidence for a saturation in 
angular momentum loss at high rotation rates (MacGregor \& Brenner 1991).
Recently Ivanova \& Taam (2003) used this method to demonstrate that
an alternate functional form for angular momentum loss rate that rises more steeply
with increasing rotation rate is consistent with X-ray data. 
This is a valuable test for the idea that magnetic field can not 
increase indefinitely with the increased rotation rate.

Using the magnetic wind model of Mestel \& Spruit (1987),
they suggested a braking law in a different form.
Based on X-ray luminosity data for fast rotators from Pizzolato et al. (2003),
they adopted the following functional dependence;
\begin{equation}
\frac{dJ}{dt}\sim-
\left\{
\begin{array}{ll}
\omega^3                & {\rm for} \; \omega \leq \omega_{crit} \\
\omega^{1.3} \omega^{1.7}_{crit}  & {\rm for} \; \omega > \omega_{crit}
\end{array}
\right.
\end{equation}

To test this prescription they compared 
the prediction of their model to the rotational velocities of stars in the Pleiades,
the Hyades and the sun. 
Given the assumed values for the initial rotation rate of a solar mass star
(close to break-up on ZAMS) and a Pleiades age of 70 Myrs, they claimed
that their braking prescription provides better fit to the data than the prescription
used in APS. Such an angular momentum loss formula would predict time 
averaged mass accretion rates an order of magnitude higher than which we could
expect from saturated law, but lower than the unsaturated prescription by a comparable factor.

In this section we demonstrate that it is difficult to match the suggested braking law 
with the value of maximum rotational velocity of solar mass stars in open clusters 
of different age. We used data on rotations of stars in four clusters (instead of 
the two used by Ivanova \& Taam 2003). Furthermore, we assumed a much more realistic 
initial rotation rate, close to the fastest rotating ZAMS stars and the initial 
angular momentum from protostars as discusses by Tinker et al. (2002). 
In addition we adopted more recent estimate for the age of the Pleiades of 130 Myrs
(Stauffer et al. 1998, Martin et al. 1998). 

The equatorial rotational velocity of a single star with solar mass and composition
as a function of age for three different prescriptions is shown on the figure 6.
The initial pre-MS star was assumed to have a rotation period of 3 days, which 
would produce a ZAMS star close to the fastest observed rotators and which 
corresponds to the upper envelope of observed pre-MS rotation rates.
Starting in the pre-MS, a star is allowed to spin up as it shrinks and 
spin down as it loses angular momentum. All magnetic braking laws were calibrated
to produce the solar rotation rate at the age of the sun. The data points 
are maximum observed rotation velocities for solar mass stars in 4 
different clusters; the Hyades (age $\sim$ 600 Myrs),
the Pleiades (age $\sim$ 130 Myrs), $\alpha$ $Persei$ (age $\sim$ 60 Myrs), and 
combined data for the young clusters IC2391 and IC2602 (age $\sim$ 30 Myrs).
The opened data point denote the age of the Pleiades of 70 Myrs, used 
by Ivanova \& Taam (2003), which was used until recently in many spindown studies.
However, recent brown dwarves lithium age estimates require an older age
of about 120-130 Myr (Stauffer et al. 1998, Martin et al. 1998).

While the prescription in form (6) works for the stated by Ivanova \& Taam (2003)
assumptions, it can be seen that suggested braking law provides a worse fit to the data 
when all of the data is accounted for. If we decrease the saturation threshold for a 
solar mass star from $10\omega_\odot$ to $6\omega_\odot$ the prescription seems able 
to reproduce the angular momentum evolution for young clusters, significantly 
overestimating the rotation rate for stars of age of the Hyades. 

It has been shown that it is important to use all available data and proper
initial conditions to constrain the spindown properties of fast rotators. 
However, even if we assume that functional dependence on $\omega$ in the form (6) is correct
and derive apropriate saturation threshold, the torque predicted by (6) would not 
be far away from the one predicted by saturated braking in the range of rotations 
at which CVs exist.
\begin{figure}[t]
\plotone{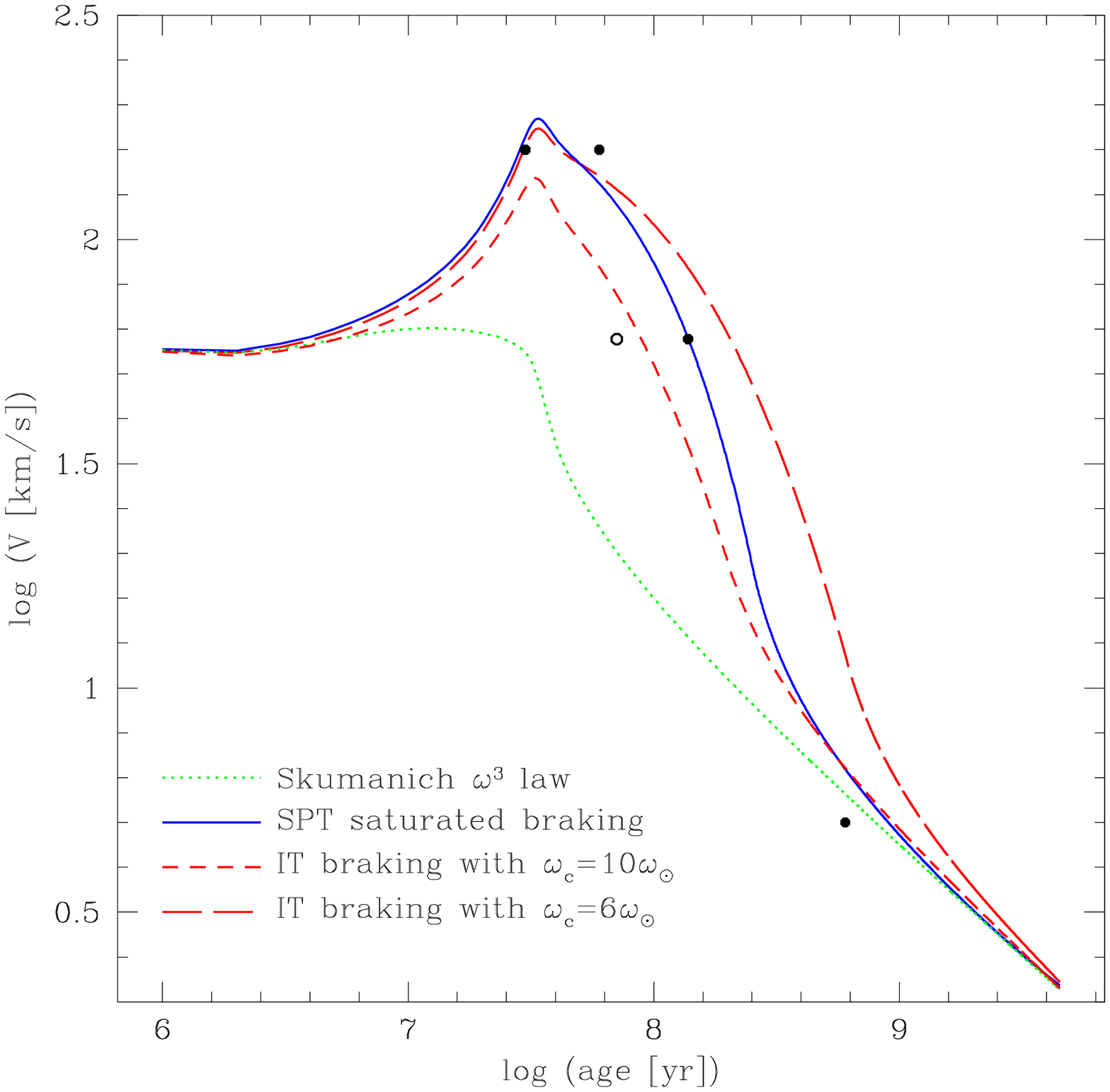}
\caption{Surface velocity of a star with solar mass and composition as a function of age
for 3 different angular momentum loss laws. Models were started at the D-burning 
birthline with $P_0=3$ days were evolved to the age and equatorial rotation period of the sun.}
\end{figure}
It should be noted, that X-ray luminosity of fast rotators provides independent 
and valuable measure for averaged generated magnetic fields, and therefore could be
a potentially very promising way to constrain the properties of the dynamo and magnetic 
braking. 

\subsection{Is magnetic braking different in single stars and CVs?}

One of the assumptions in our approach is that the empirical angular momentum 
loss rates derived for single stars can be applied to close binaries.
We assume that the gravitational field of a close companion does not affect the 
dynamo or the internal structure of the star and that it does not affect the properties 
of the stellar wind.

Stars in close binaries show on average a higher level of magnetic activity
than the single stars of the same spectral type (Simon and Fekel 1987; Schrijver 
and Zwaan 1991) However, this would be expected because of the higher (on average) 
rotation rates in tidally locked binaries, which complicates the question of whether there is 
some additional mechanism which enhances 
the generated magnetic field. Basri (1987) claimed that the differences vanished when 
this was taken into account.  

There are theoretical models of enhanced dynamo activity in tidally 
locked binaries (Zaqarashvili, Javakhishvili \& Belvedere 2002 for example); 
however, there is no strong observational evidence for 
such effects.

We therefore  conclude that the question about the applicability 
of the empirical rules for single stars to the case of close binaries does not have 
an obvious answer. In particular, the saturation threshold could potentially be affected 
by the presence of a companion.  However, the clear evidence for angular momentum loss in 
fully convective stars implies that there is no good physical basis for invoking a sharp 
decrease in magnetic braking as the explanation of the CV period gap. 

\subsection {Time averaged mass accretion rate.}
The main criticism of the application of the saturated magnetic braking derived for single stars
to the case of CVs is that predicted mass accretion rates are much smaller than 
derived from bolometric luminosities of CVs (Ivanova \& Taam 2003 for example). While such
arguments are subject to large uncertainties from our limited understanding of accretion
disk physics, it is clear that appropriately measured and rescaled mass accretion rate should be
an important test for the models of evolution of CVs. The problem which is encountered 
is that there is no direct comparison between the observed and theoretical mass 
accretion rates. Even if we disregard observational uncertainties, the observed mass 
accretion rate usually represents the instantaneous value, while the theoretical 
represents the value averaged over quite long timescales
(over many cycles of nuclear outbursts) of about $10^7$ years. For obvious reasons, if the mass
accretion rate is a variable function of time, the averaged mass accretion rate should be different 
from the instant one. Therefore, it is important to measure the 'time averaged' accretion
rate. 

Recently Townsley \& Bildsten (2003a) devised
a novel way to measure time averaged mass accretion rates in Dwarf Novas, 
determining the effective temperature of a white dwarf. Applying this method 
to about 30 DNs (Townsley \& Bildsten 2003b) they found that 
above the period gap ``unsaturated'' models overestimate the
time averaged mass accretion rate, and slightly underestimate it below the gap.

Figure 7 shows the comparison of their observations to our derived time averaged 
mass accretion rates. The observed mass accretion rate per unit surface was converted 
to time averaged mass accretion 
rate for 3 different white dwarf masses (0.6, 0.85, 1.1 $M_{\odot}$). We ran models with an initial
secondary mass of 0.9$M_{\odot}$,both from the ZAMS and pre-evolved to a central hydrogen 
abundance of $X_c$ = 0.1. We used both an unsaturated (RVJ) prescription and saturated (APS) one. 

The first conclusion is that if we assume the average mass of the white dwarf is 0.6 $M_{\odot}$, 
the predicted mass accretion rate below the period gap would not match the data for any 
magnetic braking or evolutionary state of the secondary assumed. If we increase the white 
dwarf mass, the match with data becomes much better. This in accord with the conclusion 
of Paterson (1984) that white dwarfs in CVs on average have a higher mass than 
single CO white dwarfs (around 0.6$M_{\odot}$). This most probably a manifestation 
of the first episode of 
accretion when the secondary overfills the Roche lobe for the first time and has a mass 
too large to accrete stably onto a white dwarf. In this case the stellar envelope 
expands as a result of accretion. Such accretion happens 
on a dynamical timescale, resulting in an accumulation of matter on a white dwarf 
and therefore an increase of its mass.

The second and more important conclusion is that for assumed masses of white dwarfs 
higher than those in the field, the data lies in between the derived time 
averaged mass accretion rate for both prescriptions
for magnetic braking. Therefore, it is incorrect to say that saturated magnetic braking law provides
a worse fit to the data for mass accretion than unsaturated braking law does.

Another important conclusion is that if we assume that observed mass accretion rate 
represents the close to actual values with correctly calculated uncertainties, then 
the mass loss rates observed would be insufficient to drive the secondary star out 
of thermal equilibrium and therefore this would kill
the purpose of introduction extremely efficient RVJ braking to produce the period gap. 
Ivanova \& Taam (2003) who recently suggested different form of magnetic braking 
based on X-ray activity of young stars came to the same conclusion. This would also 
imply intermediate torque for a given period
between predicted by saturated and unsaturated prescriptions.

The final thing to notice in this section is that measured mass accretion rate is averaged over
the thermal time of radiative envelope of the white dwarf ($\approx 10^3$ years) while we compare
it to the theoretical estimates of mass accretion rates averaged over many cycles of nuclear nova 
ejections ($\approx 10^7$ years). If we assume that mass accretion happens in duty cycles it would 
make the match between observations and saturated models much better, at the same time moving
observed values from the derived mass loss rates for RVJ prescription. So in this sense saturated 
prescription is preferable to provide a match between the theory and observations of 
mass accretion rates.

\begin{figure}[t]
\plotone{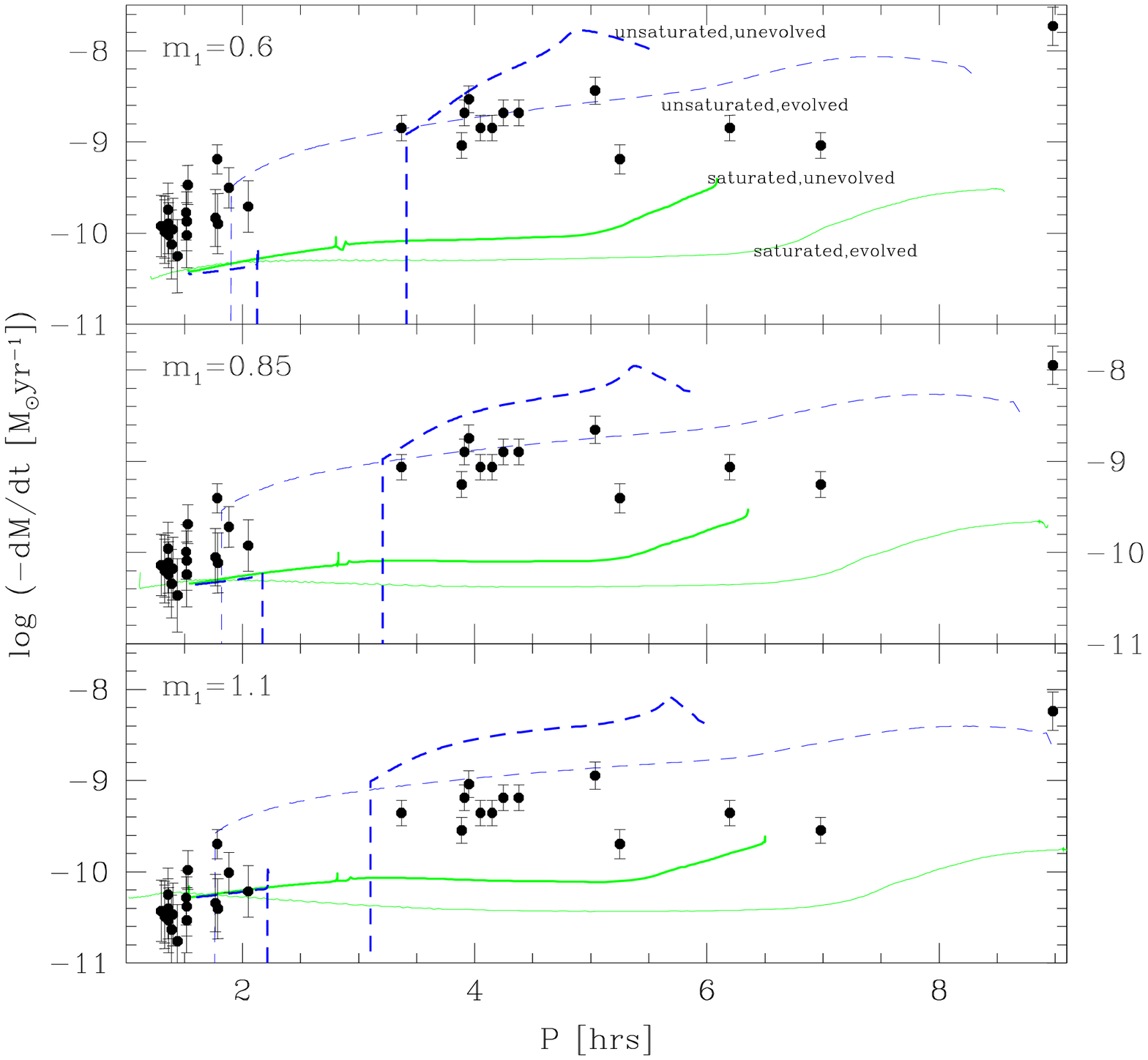}
\caption{Mass accretion rate as a function of period. Data were taken 
from Townsley \& Bildsten (2003b).
The three panels show the derived mass accretion rate for white dwarf masses of 0.6, 0.85 and 1.1
respectively, given the obtained mass accretion rate per unit surface area. Four different 
lines represent models with 
unsaturated and saturated prescriptions with secondary star on ZAMS or pre-evolved to 
central hydrogen abundance $X_c$ = 0.1. In all cases an initial secondary mass of 0.9 
$M_\odot$ was used.}
\end{figure}

\section{Summary, conclusions and speculations.}
\subsection{Summary.}
We have used the stellar evolutionary code to calculate full models of the secondary 
stars in CVs, in the attempt to understand how different prescriptions 
for magnetic braking and the evolutionary state of the secondary result in the 
observational properties of CV populations. The main features that we have focused on 
are the mass accretion rate, the mass-period relationship, and the properties 
of the 'period gap' in the distribution of CVs. Our main results are:\\
1. All models are extremely sensitive to the evolutionary state of the secondary
and less sensitive to the mass of the white dwarf.\\
2. The models in which saturated prescription for magnetic braking dictated by the 
data on spin down of single stars was used, do not reproduce the period gap. 
Models which are close to ZAMS show a shrinkage at the periods corresponding 
to transition to fully convective star, associated with the sudden mixing of 
$He^3$. This feature is not broad enough (about 0.1 hours)to produce the period 
gap. Models with evolved secondaries do not have this feature.\\
3. The saturated prescription for braking is able to reproduce the size and position 
of the gap, but only for unevolved secondaries. It fails to produce a well-defined 
period gap for highly evolved systems. It is unclear to us whether CE physics
is reliable enough and synthesis models work in detail to make predictions on the effect
of evolved models on the population of CVs.\\
4. The data for mass accretion rate (Townsley \& Bildsten 2003b) lies in between the time 
averaged mass accretion rates derived for saturated and unsaturated prescriptions. It is 
impossible to prefer one model over the other based only on this comparison with observed 
accretion rates. In addition if we assume that there is a duty cycle of accretion, 
the match with saturated braking could be better.\\
5. The everage mass of the white dwarves in CVs is required to be higher than that for single
white dwarves to match the mass accretion rate in CVs below the period gap for any prescription 
for magnetic braking.\\ 
6. Both prescriptions are able to reproduce the spread in the period-mass relation 
if evolved secondaries are included. However, for periods between 3 - 4 hours the 
scatter in mass for a given period is considerably narrower for the unsaturated prescription, 
and therefore better statistics with smaller uncertainties in this range can be a 
good potential test of the braking law for CVs. Podsiadlowski et al. (2003) predict 
that unevolved stars dominate the CV population from 3 - 4.7 hours; with sufficient
statistics this is testable. If Podsiadlowski et al (2003) are correct, one 
would therefore expect stars just above the period gap to cluster around 
the unevolved M-P relationship.\\
7. We gave an order of magitude estimates of the effects not included into recent generation 
of models; these are mixing associated with tidal distortion of the secondary star and
effect of magnetic spots on the radius. We found that these effects might be important and
therefore should be included in more sophisticated calculations.

Given the results above, it becomes inevitable that the danger of elevating the 
hypothesis of disrupted magnetic braking to a conclusion is quite real.
Therefore, it seems important to look for other possible mechanisms of formation 
of the period gap instead of ``disrupted magnetic braking model''. We summarize 
possible scenarios below. The detailed quantitative investigation of these 
possibilities, however, is a subject of some other work.

\subsection{Speculations about the origin of the period gap.}
In this section we explore the following categories of solutions:\\
1. A non-equilibrium period distribution (finite age effect).\\
2. Changes in the mass-radius relation, possibly due to effects of spots and tidal mixing.\\
3. Different populations of CVs.

\subsubsection {Finite age effect.}
The timescale for CV evolution becomes long, comparable to the age of the Galaxy if 
the angular momentum loss rates inferred from young single stars are applicable for
CVs (see figure 1).  For example, a 0.9 $M_{\odot}$ secondary goes from an initial 
period of 6.5 hr to 1.5 hr in 9 Gyrs.  

This opens up the possibility for the formation of a Period Gap which is usually
not considered; the 'finite' age effect. The distribution of binaries would not 
be steady state, and characteristics of it should be considerably dominated by the 
dependence of injection rate on time.

This idea has support from recent research on the star formation history in the 
local disk. Majewski (1993) summarizes the efforts to derive star formation rate 
(SFR) in the disk from ages for a volume limited sample of F-G stars (Barry 1988), 
white dwarf luminosity function (Noh \& Scalo 1990), and the frequency distribution 
of lithium abundances in red giants (Brown et al. 1989). He concludes that there is 
evidence for SFR fluctuations of an order of magnitude and identifies 3 major bursts 
of star formation, separated by quiescent phases;\\
1. A star formation epoch from 11 to 7 Gyr ago, (SF burst C.)\\
2. A star formation epoch from 6 to 3 years ago, (burst B.)\\
3. A recently ongoing burst of star formation, started from 2 Gyr ago, (burst A.)\\
Naturally, these variations of the SFR should have their imprint on the period 
distribution of CVs if their evolution is really as slow as predicted by th SPT 
braking law. In this case it would be logical to identify the gap as the quiescent 
phase in star formation (between burst A and B, or B and C). Moreover, if we tend to 
interpret the slight increase in the number of systems with periods about 7 hours from
a uniformly declining tail of CVs above the period gap' as a feature associated with
burst A, then it would be logical to conclude that the gap is formed as result of
quiescent SFR between burst B and C. Therefore all 3 spikes in the period distribution 
of CVs (at 2,3,7 hours) could be identified as direct fingerprints of SFR bursts C,B
and A respectively. However, it would be difficult to precisely test this idea because of the
many factors contributing to CV population.

The test for this hypothesis would be improved statistics for CVs in a single-age 
environment (stellar clusters) or a uniformly old population (the galactic halo). 
If the period distribution was different from that observed in the solar neighborhood, 
it would be a clear indication that the CV population was not in equilibrium.

\subsubsection{Possible Mechanisms for a Change in the Mass-Radius Relationship.}

There are two physical effects that we have not included in our models: 
1) structural changes in CV secondaries arising from high spot coverage, and
2) rotation-induced mixing in the cores of chemically evolved secondaries.

These mechanisms change mass-radius relationship and must be included
into full models of stars in close binaries (as it was shown in section 4),
they would affect the period of CV for a given mass of the secondary. 
However, these mechanisms doubtfully
responsible for the formation of the period gap, because neither of two mechanisms shows 
any abrupt behavior close to 3 hours.

\subsubsubsection{Starspots.}
There is reasonably compelling circumstantial evidence that rapidly rotating 
low mass stars (above the fully convective boundary) have larger radii than 
those predicted by standard stellar models.  The radii of fully convective 
stars are in accord with the predictions of theoretical models. This creates
the possibility that there could be a change in the mass-radius relationship
near the fully convective boundary that could partially explain - or even 
cause - the CV period gap. In this paper we have used standard stellar models;
those that begin the CV phase with significant nuclear evolution follow a 
different mass-radius relationship than unevolved models. Recently there has 
been a significant increase in the quantity and quality of fundamental data 
for lower main sequence stars.  This is largely due to surveys that have 
discovered several eclipsing binary systems (Ribas, 2003;
Torres \& Ribas, 2002) and interferometric radius 
measurements (S\'egransan et al. 2003). The radii of eclipsing binaries is
observed to be close to theoretical expectations for the lowest mass 
stars ($M < 0.3 M_{\odot}$) and for higher mass stars ($M > 0.8 M_{\odot}$).  However,
the radii of intermediate mass stars appears to be systematically larger
than predicted by stellar interiors models (S\'egransan et al. 2003).

Strong observational selection effects favor the detection of tidally 
synchronized eclipsing binaries, which rotate significantly faster than 
typical low mass field stars. There is intriguing evidence that rapidly 
rotating lower main sequence stars have large starspot covering factors;
these have recently been implicated as the cause of anomalies in 
photometric color-magnitude diagrams in young open clusters (Stauffer et
al. 2003). In the presence of large covering factors, the radii of the models could be affected.  
This problem was explored by Spruit \& Weiss (1986), and the results of 
their exploratory calculations are in the right sense to explain the observed 
trends.  For stars with deep convective envelopes and radiative cores the 
luminosity is insensitive to the boundary conditions, and large cool spots 
will tend to increase the radius while holding the luminosity nearly constant. 
For fully convective stars, by contrast, changes in the surface layers will 
have a direct impact on the central temperature; spots will tend to lower 
the mean surface effective temperature and luminosity, leaving the radii 
almost unaffected.

The relative change in radius and mass of a star with spot coverage 50\% is
shown on the figure 8. Spots are assumed to be completely black. This sets an upper limit
on the fractional change of stellar parameters; real spots which would have some 
some radiative energy transport through them would have less impact on a star. The change of 
radius is positive compared to an unspotted star, while the change in luminosity is negative.
While change in radius can be significant (up to 15\%) for stars with masses 1.0 - 1.2 $M_{\odot}$,
this mechanism does not produce any abrupt transition in the interesting region 
0.2 - 0.4 $M_{\odot}$. In addition it is very small in this region (1 - 2\%). 

\begin{figure}[t]
\plotone{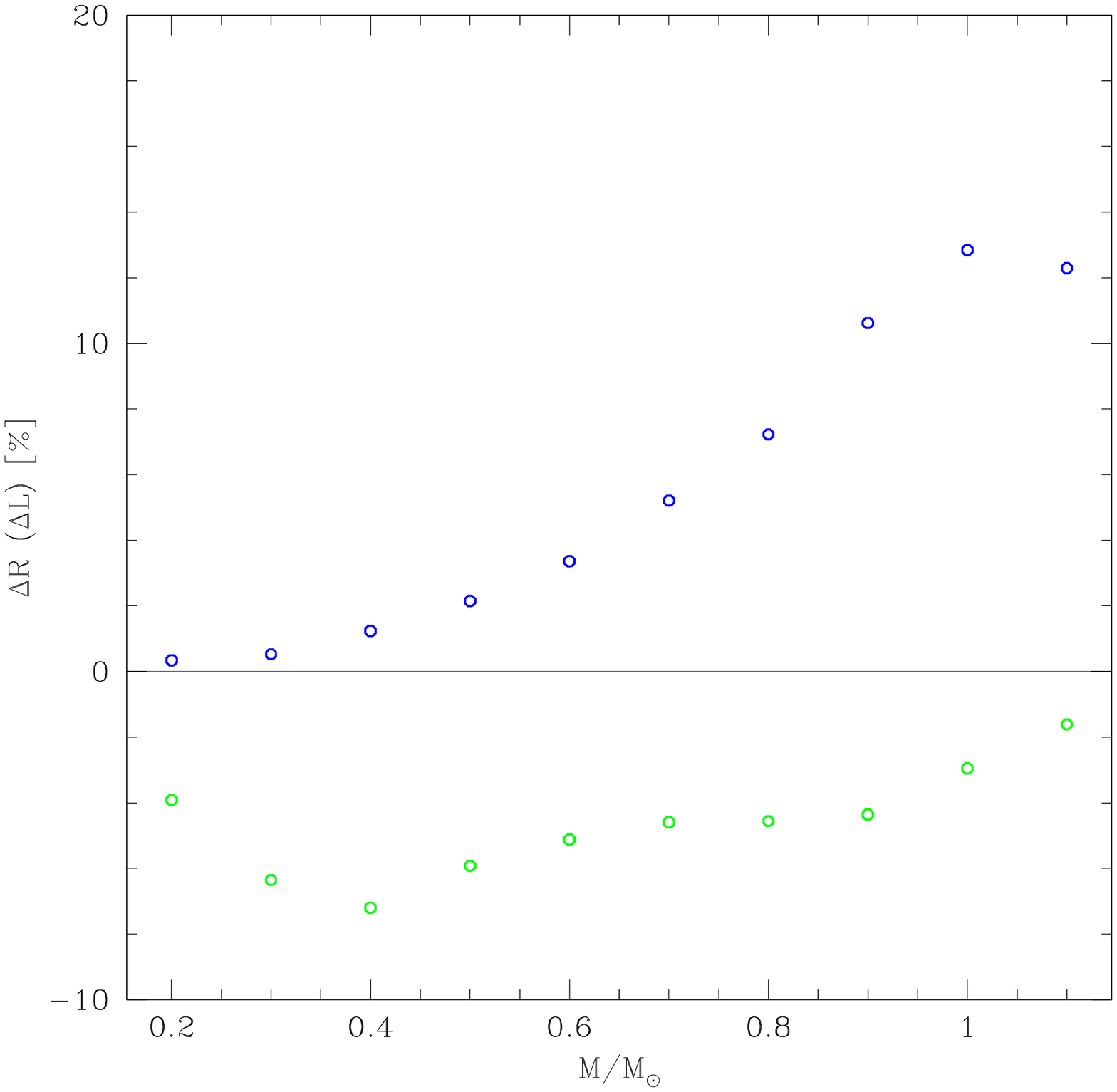}
\caption{Relative change in radius and luminosity of a star with 
effective spot coverage 50\% as a function of mass. Spots are assumed to be completely black
(see text).}
\end{figure}

\subsubsubsection{Rotational mixing.}
The second potential mechanism for changing the mass-radius relationship is 
rotational mixing.  Our main result for chemically evolved secondaries in 
CVs is that they both follow a different mass-radius relationship than 
chemically unevolved stars and that they become fully convective at a 
different mass.  Once the stars are fully convective all of the tracks 
converge on a narrow range of radii at a given mass; modest differences in 
the envelope helium abundance have little impact on the radius. These models, 
however, do not consider rotational mixing in the radiative core. Such mixing
could be driven by the internal rotation of the secondary (see Pinsonneault 
1997 for a discussion of some of the physical mechanisms) or it
could be induced by tidal distortions from spherical symmetry. The 
timescale for mixing will depend sensitively on the rotational period of the 
system; if the timescale for mixing could drop below the 
timescale for angular momentum loss at a critical period, this would cause
a transition from a large range of radii for a given secondary mass to a 
small one, as the evolved stars tend towards the radii appropriate for the 
single stars. One complicating theoretical uncertainty is the degree to 
which mean molecular weight gradients inhibit mixing.
The appropriate timescales for 3 different models are shown in the figure 9. 
\begin{figure}[t]
\plotone{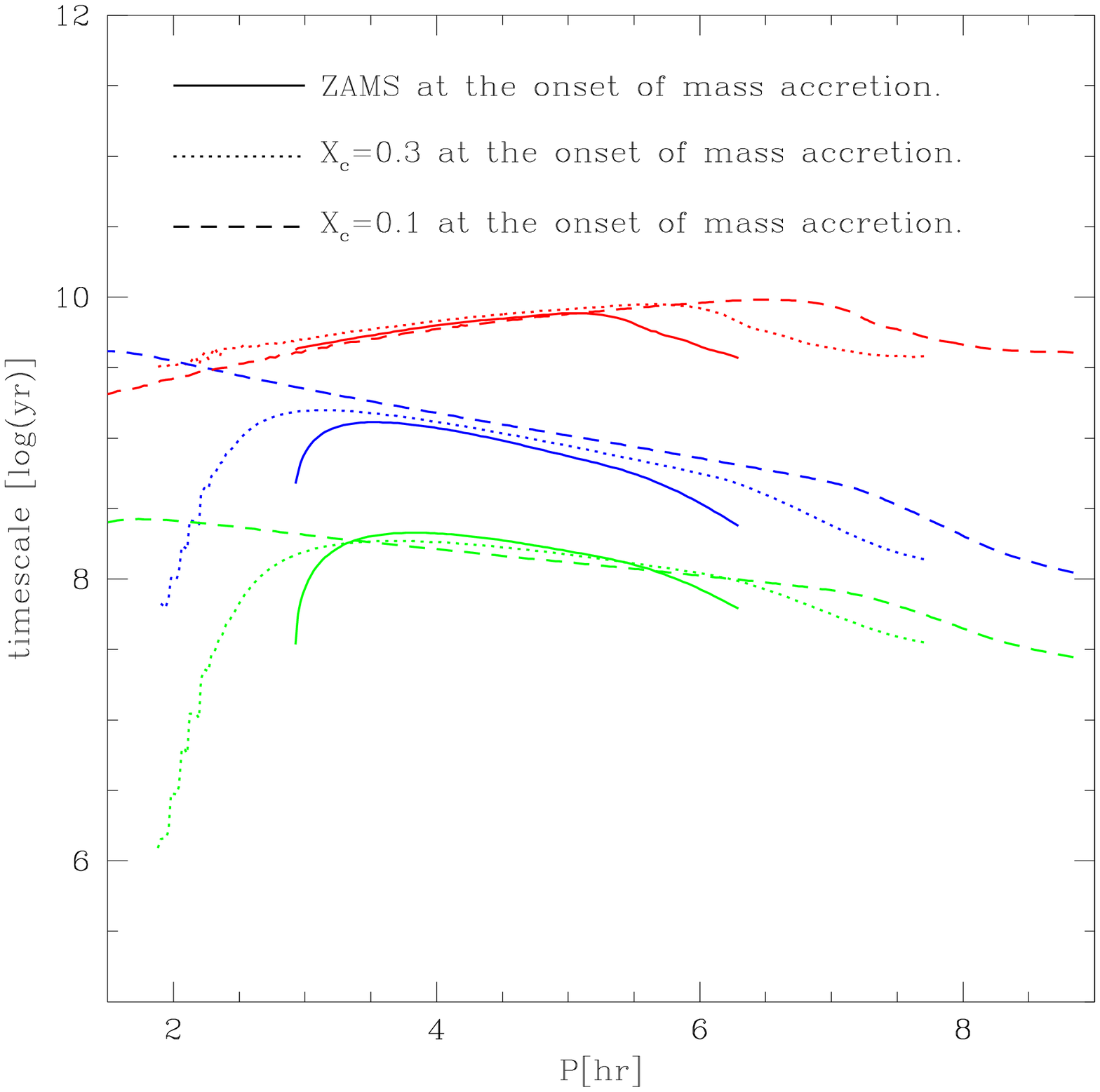}
\caption{Timescales as a function of orbital period for 3 models with 
different secondary evolutionary
states at the onset of accretion. The mass of the primary is 0.85 $M_{\odot}$, 
and the initial secondary mass
is 0.9 $M_{\odot}$. The solid lines - ZAMS model, doted lines represent model with 
initial $X_c=0.3$ and dashed lines model with initial $X_c=0.1$.
The upper curves are thermal timescales 
of radiative core for all three models as a function of rotational period. Curves in 
the middle represent the minimum timescale for the classical Eddington-Sweet 
meridional currents at the base of the convection zone. 
The bottom lines represent the minimum rotational mixing 
timescale including tidal distortion effect.}
\end{figure}

The models in figure 9 have central hydrogen abundance at the onset of mass 
accretion ZAMS value of 0.7, 0.3, 0.1 respectively. The upper curves are the thermal timescales 
of the radiative core for all three models as a function of rotational period. Curves in 
the middle represent the minimum timescale for meridional currents in the radiative interior 
for the stars rotating at this period. 

The timescale estimates above represent only the minimum possible values;
there are two important inhibiting effects. Mean molecular weight
gradients can inhibit rotational mixing (Mestel 1953; Maeder \& Zahn 1998)
by developing a latitude-dependent $\mu$ profile. Horizontal turbulence
arising from latitudinal differential rotation will tend to homogenize
level surfaces, and the net impact of $\mu$ gradients will depend on the
(uncertain) balance between the two. Second, even in the absence of $\mu$
gradients horizontal turbulence will tend to decrease the efficiency of
rotational mixing relative to angular momentum transport. This effect can
be considerable; Pinsonneault et al. (1989) found that the diffusion
coefficeints for composition mixing in the Sun were $\sim 30$  times smaller than
those for angular momentum transport (see Chaboyer \& Zahn 1992 for a
theoretical exploration of this issue). In the Zahn (1992) framework, this
efficiency factor is a function of position within a stellar
model. Exploring these effects is beyond the scope of the current paper,
and they are the subject of another paper in preparation.
However, it is clear that rotational mixing can occur on an interesting
timescale for evolved secondaries.

\subsubsection {Different populations of CVs.}
Another possibility for the origin of a period gap is that there are different populations 
of CVs which have different equilibrium period distributions.  A bimodal distribution 
could be produced if there were two distinct populations.
Some of the possible sources of such effects could include:

1. Populations might be separated by white dwarf masses (this was first raised as a possibility 
by Webbink (1979) and described and compared to others by Verbunt (1984)).  This two 
population in this scenario are He and CO white dwarf primaries. CVs are systems that 
form from binaries that undergo common envelope evolution.  They should have a 
considerable fraction of He white dwarfs - these form from systems in which runaway 
accretion from the expanding primary onto the secondary happens when the primary is on 
the red giant branch (see De Kool 1990 for an example). However, He white dwarfs have 
a considerably lower mass than CO ones (about 0.35 - 0.4 $M_{\odot}$ for the former, 
and more than 0.6 $M_{\odot}$ for the latter). As a result, the initially lower 
mass He WDs should accrete matter in a runaway process until the mass ratio of the 
primary to the secondary is high enough to allow stable accretion. This initial 
phase of accretion should increase the mass of the white dwarf, making subsequent CV 
evolution quite indistinguishable from that of a CV with an initially CO white dwarf.  
In addition there is no obvious WD mass - CV period correlation (e.g. Paterson 1984).  
Therefore, although different initial WD masses are probably present, it is doubtful 
that they can explain the period gap by themselves.

2. Populations could be separated by the evolutionary state of the secondary. The systems with 
considerably evolved secondaries on average tend to have larger orbital periods.
The larger radii of evolved systems combined with a possible change in the M-R relation 
(close to ZAMS) at the low mass end (due to spots or mixing) can be a potential mechanism 
for a formation of period gap.

3. Populations separated by the output of the common envelope phase. This is reminiscent 
of (1) with the difference that instead of having the mass of the white dwarfs which 
divide all CVs into distinct populations, the mass of the secondary and its orbital 
period serve this purpose.  If the systems with low initial secondary mass (0.1 - 0.3 
$M_{\odot}$) are close to contact after the CE phase they become CVs almost 
immediately. The systems with the intermediate secondary masses (0.3 - 0.8 $M_{\odot}$) 
will be farther away on average from the white dwarf than low mass secondaries. 
This occurs because part of the initial binding energy of a binary goes to expel 
the envelope; therefore, more massive secondaries do not need to move too close to 
a white dwarf to expel the envelope (e.g. De Kool 1990). A combination of 
larger separations with inefficient angular momentum loss can lead to 
the situation when characteristic timescale between CE and CV phases exceeds the 
age of the Galaxy. For even more massive secondaries (0.8 - 1.2 $M_{\odot}$) the 
separation is even larger. But at the same time, the saturation frequency for 
magnetic braking increases exponentially with mass; therefore, systems with higher 
mass secondaries would evolve into contact much more rapidly. It therefore might be 
possible for the CV source function to have two spikes - at lower and at higher 
secondary masses only, with few systems born that have intermediate mass 
secondaries. This could affect the period distribution in much the same fashion 
that the finite age effects do.

\section {Acknowledgment}
We would like to thank Dean M. Townsley (townsley@physics.ucsb.edu) who provided 
the data on observed mass accretion rates in DNs and 
anonymous referee for useful comments.



\newpage
\begin{references}

Andronov, N., Pinsonneault, M., Sills, A. 2003, ApJ, 582, 358


Basri, G. 1987, ApJ, 316, 377.


Baraffe, I., Kolb, U. 2000, MNRAS, 318, 354


Barry, D. C. 1988, ApJ, 334, 436


Beuermann, K., et al. 1998, A\&A, 339, 518B


Brown, J. A., Sneden, C., Lambert, D. L., Dutchover, E. 1989, ApJ Suppl, 71, 293


Chaboyer, B., \& Zahn, J.-P. 1992, A\&A, 253, 173C


Charbonneau, P., \& MacGregor, K.B. 1997, ApJ, 486, 502


Dantona, F., Mazzitelli, I 1982, ApJ, 260, 722D


De Kool, M. 1990, ApJ, 358, 189D


Eggleton, P. 1983, ApJ, 268, 368


Iben, I. \& M. Livio 1993, PASP, 105, 1373.


Ivanova, N., Taam, R. E. 2003, ApJ, 599, 516I


Keppens, R., MacGregor, K.B., \& Charbonneau, P. 1995, Astr.Ap., 294, 469


Krishnamurthi, A., Pinsonneault, M. H., Barnes, S., \& Sofia, S. 1997, ApJ, 480, 303


Landau, L. D., \& Lifshitz, E. M. 1962, {\it The Classical Theory of Fields},
(2nd ed: Oxford: Pergamon).


Lanza, A. F., Rodono, M., \& Rosner, R. 2000, MNRAS 314, 298 


MacGregor, K. B., Brenner, M. 1991, ApJ, 376, 204M 


MacGregor, K. B., \& Charbonneau, P. 1997, ApJ 486, 484


McDermott, P. N., Taam, R. E., 1989, ApJ, 342, 1019M 
 

Maeder, A., \& Zahn, J.-P. 1998, A\&A, 334, 1000M


Majewski, S. R. 1993, \araa, 31, 575


Mestel, L. 1953, MNRAS, 113, 716M


Mestel, L., Spruit, H. C. 1987, MNRAS, 226, 57M


Montesinos, B., Thomas, J. H., Ventura, P., \& Mazzitelli, I. 2001, MNRAS, 236, 877


Noh, H. R, \& Scalo, J. 1990, ApJ, 352, 605


Parker, E. N. 1993, ApJ, 408, 707P


Patterson, J. 1984, ApJS, 54, 443.


Pinsonneault, M. H. 1997, \araa, 35, 557


Pinsonneault, M. H., Kawaler, S. D., Sofia, S., Demarque, P. 1989, ApJ, 338, 424P


Podsiadlowski Ph., Han Z., \& Rappaport S. 2003, MNRAS, 340, 1214


Rappaport, S., Verbunt, F., \& Joss, P. C. 1983, ApJ, 275, 713


Ribas, I. 2003, A\&A, 398, 239


Schrijver, C.J., Zwaan, C. 1991, A\&A, 251, 183.


S\'egransan, D., Kervella, P., Forveille, T., \& Queloz, D. 2003, A\&A, 397, L5


Sills, A., Pinsonneault, M. H., \& Terndrup, D.M. 2000, ApJ, 540, 489


Simon, T., Fekel, F. 1987, ApJ, 316, 434


Skumanich, A. 1972, ApJ, 171, 565


Solokani, S.K., Motamen, S., \& Keppens, R. 1997, Astr.Ap., 325, 1039


Smith, D. A., \& Dhillon, V. S. 1998, MNRAS, 301, 767


Spruit \& Weiss (1986), A\&A, 166. 167S


Stauffer, J. R., et al. 1997, ApJ, 479, 776


Stauffer, J. R., et al. 2003, ApJ, in press


Stauffer, J. R., Schultz, G., Kirkpatrick, J. D. 1998, ApJ, 499L, 199S 


Torres, X. \& Ribas, I. 2002, ApJ, 567, 1140


Townsley, Dean M. \& Bildsten, Lars, 2003a, astro-ph/0306080


Townsley, Dean M. \& Bildsten, Lars, 2003b, ApJ, 596L, 227




Verbunt, F. 1984, MNRAS, 209, 227


Warner, B. 1995, {\it Cataclysmic Variable Stars}, Cambridge Astrophysics
Series, (New York: Cambridge University Press).


Webbink , R. F. 1979, White Dwarfs and Variable Degenerate Stars, IAU Colloq., 53, p.426
 

Weber, E. J., Davis, L. 1967, ApJ, 148, 217W


Zahn, J.-P. 1992, A\&A, 265, 115Z


Zaqarashvili, T., Javakhishvili G., \& Belvedere G., 2002, astro-ph/0210222

\end{references}
\end{document}